\documentclass[12pt]{article}
\usepackage{amssymb}
\usepackage{amsfonts}
\usepackage{amsmath}
\usepackage{amsthm}

\title{After Bell}

\author{Andrei Khrennikov\\International Center for Mathematical Modeling \\
in Physics, Engineering, Economics, and Cognitive Science\\
Linnaeus University, V\"axj\"o, Sweden }

\begin{document}
\maketitle

\begin{abstract}
 We analyze foundational consequences of recently announced loophole free tests of violation of Bell's inequality. We consider 
two interpretations of these remarkable experiments. By  the conventional one  ``Einstein was wrong and Bohr was right,  
there is spooky  action at a distance, quantum realism  is incompatible with  locality.'' However, we show that it is still possible 
to treat quantum mechanics without appealing to nonlocality or denying realism. We hope that this note will attract attention 
of the experts in quantum foundations and convince them to come with their own comments on the final Bell's test. 
\end{abstract}

\section{Introduction}

It finally happened! In 2015 three world's leading experimental groups working on foundational aspects of quantum mechanics (QM) announced (practically simultaneously) 
that they had performed  the loophole free tests of violation of the Bell inequality \cite{Bell-EPR}, \cite{B}:  the groups  of Ronald Hanson (Delf University of Technology)  \cite{Delft},  
Anton Zeilinger (University of Vienna) \cite{Vienna}
and Linden Shalm (NIST, Boulder) \cite{Boulder}. This is definitely a great event in quantum foundations; the more so because it took so long to evolve from the pioneer
experiment of Alain Aspect \cite{Aspect0} to these final Bell's tests.  (Besides, some experts 
in quantum foundations presented the arguments, in the spirit of Heisenberg uncertainty principle, that the locality and detection loopholes could not be closed in one experiment \cite{KHRV}).

I was surprised to see that this event did not  generate a new wave of the  quantum foundational enthusiasm, neither in the quantum community nor 
in general  mass-media (even oriented to popularization of science).\footnote{The situation reminds me of what happened in mathematics after the famous Fermat theorem 
had been finally proven. Mathematicians were definitely not able to ``sell'' this great achievement, even in the scientific community; the test-question: who did prove the Fermat theorem?}
One of the reasons for this rather mild reaction is that, as was already mentioned, the result was commonly expected.\footnote{One very famous experimenter whose group 
at that time worked hard to perform a loophole free Bell's test told me that he would be really excited if in this test the Bell equality would not be violated...}   A similar reason 
is that the Bell inequality has already been widely ``sold''. In literature and  talks the Bell argument is typically presented as if everything has already been experimentally proven; scientists 
working on closing loopholes were considered merely as polishing the famous Aspect's experiment  \cite{Aspect0}. Those who are closer  to quantum foundations could 
additionally point  that  G. Weihs \cite{Weihs0} contributed to complete Aspect's experiment by closing the locality loophole.  

 Therefore, it would be great if the recent publications  \cite{Delft}-\cite{Boulder} ignited a serious discussion on the possible impact of this event 
of the realization of the totally loophole free Bell test.  Such a discussion is especially important because 
conclusions presented in \cite{Delft}-\cite{Boulder}, see also comments on these tests in \cite{aaview},  \cite{Wiseman},  present only 
a part of the wide spectrum of views on Bell's argument.  Although the presented ``conventional viewpoint'' dominates in the quantum community, it would be natural to represent other,
so to say, singular, parts of this spectrum of viewpoints, see, e.g., the recent comment of  M.  Kupczynski \cite{KP3}.   

We briefly remind the conventional viewpoint presented in \cite{Bell-EPR}, \cite{B} and in hundreds of articles and monographs, 
e.g.,  \cite{CH1}-\cite{CH3}, \cite{Delft}-\cite{Wiseman}. It was finally confirmed experimentally  that 
\begin{enumerate}
\item CV1: {\it Einstein was wrong and Bohr was right;}

\item CV2: {\it  there is spooky  action at a distance; }

\item  CV3: {\it  quantum realism  is incompatible with  locality.}    

\end{enumerate}
The views of those who disagree  with the presented ``conventional position'' are characterized by the high degree of diversity  \cite{PIT}-\cite{Theo}. Therefore I shall not try 
to elaborate on some common ``non-conventional viewpoints'', but present only my own position \cite{KHR_CONT}:    
\begin{enumerate}
\item  NCV1: {\it  both Einstein and Bohr were  right;}
\item NCV2: {\it  there is  no need in  spooky  action at a distance; }
\item  NCV3: {\it  quantum realism  is compatible with  locality.}    
\end{enumerate}

  In section 2 I shall  question the CV1-CV3 viewpoint and try to justify   the NCV1-NCV3 viewpoint; in particular, I confront the ``action at a distance interpretation'' with 
the Copenhagen interpretation of QM. (It is surprising that one may combine without cognitive dissonance these two interpretations.) Then to discuss the issue of realism in QM
I appeal to its ontic-epistemic analysis in the spirit of Atmanspacher and Primas \cite{ATM}. From this viewpoint, Bell's argument can be treated as the conjecture that ontic states can be identified
with epistemic ones. We also discuss this conjecture by appealing to the old Bild conception (Hertz, Boltzman, Schr\"odinger) about the two descriptive levels of nature, theoretical and 
observational, see, e.g., \cite{SH} and chapter 1 of monograph \cite{Beyond}. Our conclusion is that the rejection of Bell's conjecture as the result of the recent experiments cannot be treated as the impossibility to keep the realist viewpoint.   
There is neither need in action at a distance. 
 
Section 3 is started with the  presentation of  Kolmogorov's interpretation \cite{K} of classical probability (CP)  as the observational theory (describing the epistemic states of nature).
By Kolmogorov CP is a contextual theory assigning probability spaces to experimental contexts, complexes of experimental physical conditions. This position leads to 
the contextual representation  of the probabilistic  structure of Bell's experimental test \cite{KHR_ENTROPY}, \cite{KHR_CONT}.

In section 4  I present my personal picture of future development of quantum foundations, in ``after Bell epoch'': from the total rejection of Bell's conjecture to novel studies on 
the two descriptive levels approach to QM. In contrast to rather common opinion (see, e.g., Aspect's paper \cite{aaview} entitled 
``Closing the door on Einstein and Bohr's quantum debate''  and Wiseman's paper \cite{Wiseman} entitled   ``Quantum physics: Death by experiment for local realism''), 
for me the final Bell test did not imply the total impossibility to ``go beyond quantum'' \cite{Beyond}. The main 
message of this test is that the way to a proper subquantum model is more tricky than it was hypothesized by Bell. \footnote{ I also remark that even those who would not be convinced 
by my philosophic and probabilistic arguments should agree that we have no choice - we have to go beyond quantum, at least to solve the problem of merging 
of two great physical theories: QM (or more precisely QFT) and general relativity.}

\section{Schr\"odinger, Einstein, Bohr, and Bell: philosophy  meets quantums physics}
\label{RE}

Typically the output of Bell's argument (nowadays experimentally confirmed) is formulated as {\it QM is incompatible with local realism.} Thus either one has to reject the possibility 
of the realist description of quantum phenomena or to imagine nonlocal reality of the Bohmian type  which is even more exotic and mystical than quantum surreality.
Since Einstein dreamed for the realist interpretation of quantum phenomena and he considered nonlocality as an {\it ``absurd alternative''} to realism,  
nowadays it is clear  (for everybody, besides a few outsiders, see, e.g., \cite{Beyond}, \cite{Hess}, \cite{KP2}-\cite{Theo}, \cite{KP3}) 
that he was wrong and automatically (as Einstein's opponent) Bohr was right,
see, e.g., \cite{aaview}. 
In general in the modern quantum community it is fashionable to be a follower of Bohr.  

\subsection{Bohr versus Bell}

First of all, I want to point to one of the main interpretational problems of the conventional viewpoint on the output of Bell's test, see CV1-CV3.  This is an attempt to present   CV1-CV3
 in a single package and 
consistently with the Copenhagen interpretation of QM.  However, none of the fathers of QM, neither Bohr, Heisenberg, Fock, Landau, Pauli nor Einstein, could even imagine
that   spooky  action at a distance would be taken seriously  in the quantum foundations. Thus, one has to recognize that the {\it  modern nonlocal feature of QM motivated 
by Bell's argument has nothing to do with Copenhagen.}  One can agree or disagree with this statement, but in any event it has to be presented and discussed.

Thus to be consistent one has to speak about {\it the revolutionary  changes in quantum foundations generated by Bell's argument} and experimental tests confirming it.
The possibility of these changes was not even imagined by earlier Copenhagenists.

One can still argue that ``Copenhagenists were against realism'' and, hence, Bell's test can be considered as supporting the Copenhagen position...

\subsection{Realism: philosophy meets physics}

Realism is a complex issue, both in physics and philosophy. One cannot debate it without taking into account the recent progress in philosophical studies.
And I want to analyze Bell's argument in the light of such studies, namely, {\it the ontic-epistemic viewpoint} on physical reality, firmly established 
in quantum foundations by H. Atmanspacher and Primas \cite{ATM}.

There are ontic states, assigned to physical systems as ``they are'', and epistemic states representing knowledge that observers gain from measurements
on physical systems. QM is about epistemic states. This is in complete agreement with the Copenhagen interpretation of QM, especially Bohr's views. 

It often escapes notice that the EPR-argument \cite{EPR} was about ontic states;   it was directed to show that quantum states are epistemic 
states.   For Einstein,  QM was incomplete in the sense that there should exists a finer description of physical processes in the microworld  
than given by QM. The states of such a subquantum model are ontic states. Einstein and his coauthors did not have any intention to treat 
the wave function $\psi$ as the ontic state; for them, it was the epistemic state representing knowledge which can be earn from measurements. 

This separation between the two descriptive levels is very important for understanding the real tragedy of QM: misunderstanding about the issue of completeness of QM, between
Einstein and Bohr, see \cite{EPR}, \cite{BR}.  As was pointed out,  Einstein's incompleteness is so to say {\it ontic incompleteness} of QM, but   Bohr's 
completeness is {\it epistemic completeness} of QM. Bohr in his famous reply to Einstein \cite{BR}
wanted to show that QM is complete at the epistemic level. However, Einstein did not have doubts in epistemic completeness of QM. 
On the other hand, it seems that Bohr was not so strongly against the possibility of creation of subquantum models; he simply considered 
such studies as meaningless, because for him physics was only about epistemic states.\footnote{ N. Bohr as well as W. Heisenberg were strongly influenced 
by E. Mach who advocated empiricism as the foundational principle of science.  This principle is the cornerstone of the Copenhagen interpretation of QM. 
At the same time it is not so common to refer to Mach as one of fathers of this interpretation; in particular, because his orthodox empiricism once had already 
 led to aggressive  denial of a basic and nowadays commonly recognized physical structure - the atomic structure of matter. }  (This picture of Bohr's views resulted from my long discussions 
with A. Plotnitsky, cf. \cite{PL}). However, it seems that Einstein and Bohr did not understand that their debate could be resolved in purely philosophic (ontic-epistemic) terms.

Now, let us consider Bell's project from the philosophic viewpoint, by using the ontic-epistemic approach. In philosophic  terms, {\it Bell's conjecture} 
was that 

\medskip

{\bf O=E}: {\it Ontic states of quantum systems can be identified with epistemic states.}
   
\medskip

Since the values of observables are assigned directly to the ontic states, 
\begin{equation}
\label{ZZTT}
\lambda \to a(\lambda),
\end{equation}
these states  also have to interpreted as the epistemic states.   Here $a$ is an observable and the result of a measurement is determined directly by the parameter $\lambda.$
Hence, $\lambda$ has to be considered as the epistemic state. At the same time Bell treated $\lambda$ as representing the state of reality as it is ($\lambda$ has no relation 
to context of measurement of $a).$ Hence, it has to be considered as the ontic state.     

The final loophole free Bell tests demonstrated that this conjecture was wrong (if there is no action at a distance, see NCV2).

However, this statement is not the same as to state that realism is incompatible with QM. The impossibility to identify two descriptive levels, 
ontic and epistemic, is not rejection of the possibility of the realist description.

\subsection{Impact of Bell's test for Copenhagenists}

For Bohr, only epistemic states had physical meaning. Of course, he did not use this terminology, but his statements match it very well; for example \cite{BR7}: 

\medskip  

{\footnotesize\em ``This crucial point, which was to become a main theme of the discussions reported in
  the following, implies the impossibility of any sharp separation between the behaviour of atomic
  objects and the interaction with the measuring instruments which serve to define the conditions
  under which the phenomena appear. In fact, the individuality of the typical quantum effects finds
  its proper expression in the circumstance that any attempt of subdividing the phenomena will
  demand a change in the experimental arrangement introducing new possibilities of interaction
  between objects and measuring instruments which in principle cannot be controlled. Consequently,
  evidence obtained under different experimental conditions cannot be comprehended within a single
  picture, but must be regarded as complementary in the sense that only the totality of the
  phenomena exhausts the possible information about the objects.''}

\medskip

Therefore it seems that the problem of identification of ontic and epistemic states was of no interest for Bohr and ``old Copenhagenists.''

\subsection{ Bild concept: Schr\"odinger versus Bell}

We remark that the ontic-epistemic approach can be considered as the modern version of the old {\it Bild-conception tradition} (Hertz,
Boltzmann and their followers as well as Einstein) \cite{SH}, \cite{Beyond}, chapter 1. In QM this tradition was especially strongly presented in views of Schr\"odinger who inherited this
tradition from his teacher Exner. By the Bild-conception one has to separate two descriptive levels, {\it one is theoretical and another 
is observational.} The latter can be identified with the modernly used epistemic  level. The theoretical level, as preceding the observational 
level, can be coupled to the ontic level of description. However, it seems that these notions cannot be completely identified. The theoretical 
level was not about ``reality as it is'', but about its theoretical mathematical model.

Schr\"odinger emphasized the impossibility to identify these two descriptive levels: theory and observations are not
necessarily related in a term-to-term correspondence and a certain degree of independence exists between them.
He also point out that already physicists of 19th century, e.g., Hertz, Boltzmann understood this well.\footnote{Boltzmann asserted that only one half of our
experience is ever experience. }  As was emphasized  by D'Agostino \cite{SH}, p. 351,    Schr\"odinger called ``the classical ideal of uninterrupted
continuous description'', at both observables' and theoretical
levels, an ``old way'', meaning, of course, that this ideal is no
longer attainable. He acknowledged that this problem was at the
center of the scientific debate in the Nineteenth and Twentieth
centuries as well,  \cite{SHX}, p.24:

{\footnotesize\em ``Very similar declarations...(were) made again and again by
competent physicists a long time ago, all through the Nineteenth
Century and the early days of our century...they were aware that
the desire for having a clear picture necessarily led one to
encumber it with unwarranted details,.'}

\subsection{De Broglie: hidden probabilities}

The excellent presentation of De Broglie's viewpoint on quantum measurement and its consequences for Bell's argument can be found in the paper of G. Lochak \cite{Lochak}. 
We now present this measurement theory briefly by trying to refine it from the presence of the physical guiding waves, De Broglie's waves. 

In his measurement theory De Broglie assigned the special value to the position measurement: every physical quantity can be measured only via the final detection 
of the position, e.g., localization of a detector produced a click. The unitary equivalence of various representations in the Hilbert space formalism is just a mathematical 
feature of the theory. To measure the concrete physical quantity $A,$ systems emitted by a source have to pass an analyzer, e.g., PBS, and then approach the 
corresponding detector. Probabilities determined by detectors after passing of the analyzer are called {\it the present probabilities;} probabilities before the analyzer, 
corresponding to the prepared pure state, are called {\it the predicted probabilities.} The main point of De Broglie's probabilistic considerations is that it is impossible 
to combine the predicted and present probabilities straightforwardly, i.e., by using the rules of the classical probability calculus, see section \ref{KOLM}.   In particular, it is impossible to combine 
the predicted and present probabilities corresponding to two arbitrary analyzers withing classical probability theory, by using the joint probability distribution.

However, this fact (a simple consequence of the quantum formalism leading to ``interference of probabilities'', see \cite{KHR_CONT} for details) does not contradict to a possibility 
to introduce hidden variables, because (according De Broglie) their probability distribution is hidden, i.e., we have to consider the third type of probabilities, {\it hidden 
probabilities}. The main point of De Broglie's analysis is that there is no reason to identify these hidden probabilities, probabilities for ontic states with 
probabilities produced by QM, namely, the predicted and present probabilities (which are epistemic probabilities). Thus similarly to Schr\"odinger, Einstein, Atmanspacher, Primas and the author of this paper, De Broglie and Lochak   
were against  identification of ontic and epistemic states and more generally their probability distributions (the latter is the main point of a long series 
of publications of the author of this note, starting with  \cite{KHRB98}).
 
 Finally, we mention the guiding wave element of De Broglie's picture. The physical guiding wave appears as the element of the De Broglie's ontic description of micro-phenomena,
i.e., as one of variables representing the ontic state. We emphasize that the above scheme with hidden (ontic) and observational (epistemic) probabilities can be explored in general, i.e., 
without claiming that a subquantum model has to be of the guiding wave type. Here we cannot discuss even briefly the difference in views of L. De Broglie and D. Bohm. We just remark 
that anybody who read De Broglie personally \cite{DB} would never say that De Broglie {\it double solution} model and so-called Bohmian mechanics coincide up to the degree of mathematical 
rigorousness. In particular, we stress that {\it  the double solution model is local} (as was emphasized by De Broglie \cite{DB}) and Bohmian mechanics is nonlocal.    

\subsection{Einstein and Bohr}

The above discussion about the ontic-epistemic philosophy and the corresponding structuring the notion of realism demonstrated that the Bell argument has no 
direct relation neither to the position of Bohr nor Einstein and, hence, it cannot be used to justify the position of one of them in contrast to another. In particular,
Einstein would say that he never tried to identify the ontic states with the epistemic states.
As was emphasized, Bohr and other Copenhagenists were interested  only in the epistemic states. Schr\"odinger would say that he is not surprised that the coupling between the (possible) theoretic
and observational levels is not so straightforward, see (\ref{ZZTT}),  as in the Bell-argument.   

I hope that the statement NCV1 has been completely justified: both Einstein and Bohr were right, but with respect to the corresponding levels of description.

\subsection{Bell versus von Neumann}

We remind that Bell  started his foundational project with hard critique  \cite{B} of von Neumann's theorem \cite{VN}. Nowadays the latter is practically forgotten, not least because 
Bell's critique.  However, we can ask ourselves: Was von Neumann's theorem really so bad?

Von Neumann did not identify the ontic and epistemic states. He established the special rules of correspondence between the two descriptive levels. In principle, he proceeded in the 
Bild concept tradition: there are theoretical and observational models and correspondence rules. Bell criticized von Neumann's image of a possible coupling between 
the theoretical and observational models and their critique was reasonable. In particular, by von Neumann the correspondence 
 between ontic and epistemic quantities should be additive,
\begin{equation}
\label{ZZTT1}
a+b \to \hat{a} + \hat{b},
\end{equation}
which seems to be unacceptable for incompatible observables.

However, from the philosophic viewpoint the sharp distinguishing between the two descriptive levels present 
in von Neumann theorem (personally he called this statement ``ansatz'') is preferable comparing with simple identification of these levels. 
A follower of Einstein would say \cite{KHR_CONT} that von Neumann simply was not lucky to find the right correspondence rules.   
As was emphasized in \cite{KHR_CONT},  the ``no-go'' activity in establishing the correspondence between the two levels of description is totally meaningless,
since there is an infinite variety of possibilities to establish the rules of such a correspondence, see, e.g., \cite{Beyond}. 

\subsection{``Bell was wrong'' community}

It is useful to analyze from the ontic-epistemic viewpoint not only  the position of the 
conventional quantum (and more specifically quantum information) community, see CV1-CV3, but even of ``Bell was wrong'' 
community. It seems that the majority of opponents of the CV1-CV2 interpretation do not understand the philosophic meaning of the Bell conjecture, {\bf O=E}.
Typically one from the  anti-Bell opposition  tries to show that observational (epistemic) probabilities can violate Bell's inequality, or to show that the standard proofs
of the Bell inequality do not work, or to construct an epistemic (local) model producing correlations violating Bell's inequality. In more tricky studies there is even a subquantum 
model and special rules coupling two descriptive levels, but, of course, such a correspondence is more complex than conjectured in {\bf O=E}. Unfortunately, such studies are not treated from 
the double descriptive viewpoint; in particular, the rules of correspondence between subquantum and quantum models are not specified, cf., however, with \cite{Beyond}.

 Then it is claimed that  ``Bell's theorem'' (argument) was wrong. My impression is that this was precisely the way of treatement of the Bell argument by  the main anti-Bell gurus: L. Accardi, K. Hess and W. Philipp,
M.  Kupczynski,  H. De Raedt,  K.,  Michielsen T.  Nieuwenhuizen and partially myself (although later I proceeded in the De Broglie-Lochak manner by distinguishing 
hidden probabilities form epistemic). The claim ``Bell was wrong''  acts as a red muleta used by Spanish matadors  to make bulls angry and generates the strong critical feedback
from the pro-Bell community.
This ``Bell was wrong'' position was not fruitful for quantum foundations, it induced brutal debates (in particular, during the V\"axj\"o series of annual conferences, 2001-2015).
By understanding that Bell's argument is based on the  {\bf O=E} conjecture and that activity to test Bell's inequality is directed to check this conjecture 
we would escape these debates or least to make them less brutal.   

In short, {\it it is definitely possible to construct local probabilistic models violating the Bell inequality, but for such models the Bell conjecture {\bf O=E} is clearly wrong; 
in the same way the standard proofs of the Bell inequality work only under the Bell conjecture.}

\section{Classical probability meets quantum physics}
\label{KOLM}

Bell's argument is of the probabilistic nature and therefore it is natural to analyze it from 
the viewpoint of foundations of probability theory.  The latter was formalized in the measure-theoretic  framework
by A. N. Kolmogorov \cite{K}  in 1933. As any scientific theory, it was endowed with the 
corresponding interpretation \cite{K}.
 
By Kolmogorov \cite{K}, see also \cite{KHR_CONT} for discussions, any complex of physical 
conditions $C,$ experimental context\footnote{Here, see  \cite{KHR_CONT}, the notion of ``context'' started to be in the use.
It is important to point out that here context represents ``total experimental arrangement''. Thus contextuality is understood wider than just dependence on measurement 
of a compatible observable. Our viewpoint on contextuality matches better the views of Bohr than commonly used, e.g., in discussions on Bell's argument,
``restricted contextuality''. We follow Bohr's recommendation that all experimental arrangement has to be taken into account.}, determines its own probability space 
${\cal P}_C$ - the system of events 
${\cal F}\equiv {\cal F}_C$
and the probability $P\equiv P_C.$ \footnote{The system of events has the structure of  Boolean $\sigma$-algebra and the probability 
is the $\sigma$-additive measure normalized by 1. Here $\sigma$ express countability: ${\cal F}$ is closed 
with respect to countable unions and intersections and probability is countably additive.}   

Classical probability (CP) theory was designed to serve the epistemic description. Here the positions 
of Kolmogorov and Bohr are very close.   In CP complementarity is expressed in assigning 
different probability spaces to different contexts.\footnote{
One of the problems of modern CP is that this clear contextual message of Kolmogorov is not emphasized  in presentation 
of CP in modern textbooks; thus the young generation of probabilists lost this original multi-context and multi-space picture of CP.
Typically the context-indexing is omitted. At the beginning it was done only to simplify notations and probabilists
(at least representatives of the Soviet school) remembered well Kolmogorov's slogan: {\it ``no probability unrelated to 
experimental context''.} When, say 15 years ago, I spoke with two former students of Kolmogorov, A. N. Shiryaev 
and A. V. Bulinski, and tried to teach them that CP is contextual, they immediately pointed to section 2 of Kolmogorov's book \cite{K}.  For them,
it was totally evident that each experimental context generates its own probability space. It seems that this rule is forgotten in 
the modern probabilistic community.}

However, often we have to work with probabilistic data collected for a family of contexts. Therefore CP has to provide some mechanism of embedding 
of multi-contextual data in a single probability space. It is important to emphasize that there are two main CP constructions of such embedding:
\begin{enumerate}
\item {\it Marginal embedding:} contextual probabilities  are represented as marginal probabilities for a single probability 
measure $P.$
\item {\it Conditional embedding:}
 contextual probabilities  are represented as marginal probabilities for some probability for a single probability 
measure $P.$
\end{enumerate}
 The first (straightforward)  embedding was excellently realized by A. N. Kolmogorov in his famous theorem \cite{K} about existence of the probability measure $P$ 
(defined on the space of trajectories) for a stochastic process determined by its   probability distributions for all finite sequences of instances of time,
$P_{t_1 ... t_n}.$ 

However, this construction can work only in special cases, for special families of contexts. (Therefore Kolmogorov's theorem about the existence of a probability 
space for a stochastic process is so famous: he found such a special (and very important) case.) It is well known in CP, since the work of Boole, that probability 
distributions for random variables measured for different contexts can be incompatible, i.e., it is impossible reproduce them as marginal probabilities from the 
the single probability $P$ \cite{Boole}. \footnote{In the quantum foundational community this historical fact was ``discovered'' by I. Pitowsky; in particular, he pointed to it 
in his talk at V\"axj\"o-2001 conference. For many years, I tried to advertise it for experts in quantum foundations, but without any success.} In fact, the inequality providing the necessary condition of the existence of such $P$ derived by Boole coincides with 
Bell's inequality; therefore some authors even proposed to speak about {\it the Boole-Bell inequality \cite{KHR_ENTROPY}.} The complete description of conditions of existence of the straightforward 
embedding of contextual probabilities in a single probability space was presented in the paper of Vorobj'ev \cite{VR} which contains all Bell's type inequalities 
known in quantum physics (as well as still unknown). 

We remark that the marginal embedding  is based on the identification of the 
ontic and epistemic descriptions. The epistemic probabilities for fixed contexts are identified with the marginal probabilities with respect to the probability distribution 
$P$ which can be treated as the ontic probability.  In particular, the Kolmogorov theorem on the probability space for a stochastic process can be applied only the assumption 
of the ontic-epistemic identification; without this assumption one has to use theory of quantum stochastic processes \cite{ac}. 

However, in contrast to QM, the ``no-go'' activity, the impossibility theorems about striaghtforward marginal embedding was not so popular in CP.  And it is clear why. There 
is widely explored the second embedding procedure based on the identification of concrete contextual probabilities not with marginal, but with conditional 
probabilities with respect to a single probability measure $P.$  This is so-called {\it randomization procedure.} It is especially important in statistics. 
We shall present it for the concrete multi-contextual test - the Bell test.

\subsection{Contextual probabilistic structure of Bell's test} 

The considerations of this section are presented in the form of a formal mathematical model in the article \cite{FOOP_B_KHR}.

How would Kolmogorov handle Bell's argument? For Kolmogorov,  it is based on consideration of 
a few experimental contexts $C_{\theta_i \theta_j^\prime}$ corresponding to fixing the pairs of orientations of PBSs. 
There are a few probability spaces    corresponding to these measurement 
contexts.  Experimental statistical data collected for  context 
$C_{\theta_i \theta_j^\prime}$ has to be handled in its own probability 
space  ${\cal P}_{C_{\theta_i \theta_j^\prime}}.$

\begin{itemize}
\item Can Bell's type inequalities be derived in such multi-space probabilistic framework?  Not!
\item  Would be Kolmogorov surprised by violation of Bell's inequality for multi-space data?
Not at all! Moreover, he would be surprised if it were not violated. 
\item Can one construct a single probability space representing all contexts $C_{\theta_i \theta_j^\prime}?$
Yes!  Construction  is known as  randomization.
\end{itemize}

In short randomization can be described as follows: 

\begin{itemize}
\item A).  Choose the probabilities  of selections of  contexts.
\item B).  Determine probability distributions for fixed contexts,  $P_{C_{\theta_i \theta_j^\prime}}.$
\item C).  Construct probability  $P$ serving for all experimental contexts:  combine probabilities for fixed contexts $P_{C_{\theta_i \theta_j^\prime}}$ with probabilities 
of context-selections.\footnote{ This  $P$ is defined on the system of events ${\cal F}$ related to this randomized experiment; it contains not only events related to observations
for  fixed contexts $C_{\theta_i \theta_j^\prime},$ but also events of selections of these contexts.} 
\end{itemize}
This single probability measure $P$  represents the multi-context experiment. 
We stress that,  for it,  it is possible to derive Bell's inequality, since the latter is just a theorem of CP \cite{KHR_CONT}.

In this framework the probability distributions for fixed contexts $P_{C_{\theta_i \theta_j^\prime}}$
(which are obtained by experimenters as frequencies)
appear as  conditional probabilities with respect 
to fixing selections of the parameters $\theta_i,  \theta_j^\prime$ (contexts $C_{\theta_i \theta_j^\prime}).$ 
However,  {\it  there is no reason  to hope that these conditional probabilities and corresponding conditional quantities, 
e.g., correlations, would satisfy Bell's inequality,} see \cite{FOOP_B_KHR}.
 
In CP  the magnitude $\Delta$  of  violation of Bell's inequality can be treated as the numerical measure of 
multi-space probabilistic structure of the experiment. Thus   $\Delta$ represents the degree of contextuality (which is understood a la Bohr as taking into account 
all experimental arrangement). What does such contextuality mean from the QM foundational viewpoint? Contextuality exhibits itself only the existence of incompatible contexts.
Therefore  $\Delta$ is also the measure of incompatibility-complementarity. From this viewpoint, the experimental tests violating Bell's inequality confirmed complementarity 
not  only for local observables, such as position and momentum or two spin projections of a single electron, but even for nonlocal observables such as projections of spins measured
for a pair of electrons.

\subsection{Kolmogorov versus Bell} 

As we have seen, the CP description of Bell's test does not lead to revolutionary consequences in the form CV2, CV3.
And this is completely clear why. The above CP model of the probabilistic structure of Bell's test is epistemic - it takes into account 
randomness of selection of experimental contexts, i.e., random generators used for this aim are also considered as measurement devices.
In principle, one can say that this is the end of the CP-story about Bell's test.
 
However, we can continue our discussion  and represent its output in the  hidden variable form.  The latter will be not real ontic  hidden variables, but 
so to say ``epistemic hidden variables''. 

CP teaches us that all random influences involved in the tests have to incorporated into the model.  Probabilist would consider \cite{FOOP_B_KHR} not  the``Bell's map''
(\ref{ZZTT}), but a map of the form:
\begin{equation}
\label{ZZTT2}
\omega \to a(\omega), \; \omega=(\lambda, \lambda_{L}, \lambda_{R}),   
\end{equation}
where $\lambda_{L}, \lambda_{R}$ are random parameters determining with some probabilities, $p_L, p_R$ outputs of random generators $ L$ and $R$ which in turn
determine settings.\footnote{In Bell's test for the CHSH inequality each random generator selects a pair of angles with probabilities  $p_L(i), P_R(j), i,j=1,2.$} 

Of course, not only random generators contribute to randomness generated by experimental equipment. There is a lot of randomness in PBSs, in detectors, in optical fibers. 
This randomness also has to be taken into account by modifying (\ref{ZZTT2}).

\subsection{Classical probability, operational approach and Bell's argument}

The above CP-analysis of Bell's test motivates us to comment the representation of observables in the form (\ref{ZZTT2}) from the operational viewpoint which is widely used in QM, e.g.,  \cite{DE_M}.
By it there is a preparation procedure and there is a measurement procedure. From the operational viewpoint, in Bell's considerations the preparation procedure is performed 
by the source of entangled systems  and the detection procedure by detectors. However, there are, e.g.,  random generators determining the orientations of PBSs. They exist! 
One cannot simply ignore them. In the consistent 
operational approach they have to taken into account.  In the model represented by  (\ref{ZZTT2}) they are included in the preparation procedure. In fact,  we follow
De Broglie's who emphasized the role of analyzers in  production of the present probabilities. In the Bell test for photons PBSs are precisely De Broglie's analyzers. 

\subsection{With Ockham's razor against nonlocality}
 
We emphasize that we have not appealed to spooky  action at a distance. Contextuality \footnote{We understand it as determination of probability space by 
experimental context.} leads to violation of Bell's inequality. This motivates NCV2 -  ``there is  no need in  spooky  action at a distance'',  although not rejection of CV2.  
In principle, one can still play  under the assumption  CV2 -  if one likes spooky action and it helps her to create the consistent 
picture of the world.   However, there  is no need in it. This is the good case to use Ockham's razor, see \cite{TIME} for discussion. 

This viewpoint, that by taking into account Kolmogorov probabilistic contextuality of the Bell test one need not mention nonlocality at all, is often criticized by piointing out 
that    contexts $C_{\theta_i \theta_j^\prime}$ are ``nonlocal'', since their are based on orientations of two spatially separated PBSs. However, this classical nonlocality 
of spatial location of experimental equipment has nothing to do with action at a distance.

Another point which was discussed in \cite{KHRV} is that in the Bell conjecture  the notion of ``locality'' is represented in very specific form of action at a distance \cite{B}. Where are there 
space-time variables? Action at a distance locality is not at all Einsteinian locality of relativity theory. In principle, one can try to couple ``Bellian and Einsteinian localities'' by taking into account 
the space-time dependence of correlations for entangled systems \cite{BN}. However, by proceeding this way we lose the simple prequantum-quantum (ontic-epistemic) correspondence 
given by (\ref{ZZTT}). The real space-time correlations contain the at least  contribution of media for signal propagation; then one has to take into account dispersion and losses in this media, finally 
the temporal structure of functioning of photo-detectors \cite{BN1}. Thus the model becomes epistemic; violation of Bell's inequality by such correlations is not surprising.

\subsection{Quantum realism}
\label{QR}

Bell's argument strongly supported anti-realist tendency in the quantum community. It was very supporting for development of a variety of information interpretations of QM -
with two bright examples, the Zeilinger-Brukner information interpretation \cite{ZB} and Fuchs' QBism \cite{Fuchs1}, \cite{Fuchs2}. 
Nowadays a randomly chosen representative of the quantum community 
would say with very high probability (I account it as 99\%) that {\it realism and QM are incompatible.}

I totally disagree with this position, see, e.g., my manifest \cite (may be naive, since at that time I was not so well educated in quantum foundational issues). I repeat 
that the successful Bell's test only rejects the straightforward coupling (\ref{ZZTT}) between the ontic and epistemic or more generally theoretic and observation levels.
Other maps can be constructed and the fact that we still do not have a commonly acceptable map may just mean lack of imagination, cf. with the famous citation 
of Bell \cite{B1}, p. 15: {\footnotesize\em  ``what is proved, by impossibility proofs, is lack of imagination.''}

However, realism discussed in section \ref{RE} is merely subquantum realism. What is about, so to say, genuine quantum realism?

As was shown, CP modeling of physical experiments, including multi-contextual experiments such as Bell's test, does not lead to rejection of realism. This is realism of experimental 
contexts and observations performed for these contexts. This is precisely  ``physical realism'', defined as the statement that {\footnotesize\em the goal of physics is to study entities of 
the natural world, existing independently from any particular observer's perception, and obeying universal and intelligible rules,} see Auff\`eves and Grangier \cite{GR3}.

In my works this realism was treated as {\it contextual realism} of QM \cite{KHR_CONT}. However, this terminology might be misleading, since in quantum foundations contextuality 
is treated restrictively, as dependence on outcomes of joint measurement of a compatible observable. We can also speak about epistemic realism. 

Finally, we remark that physical (contextual) realism matches well with Bohr's views.

One may even say that my contextual quantum realism (similarly quantum realism of 
Grangier)\footnote{To reconcile realism with QM, Grangier et al.  also proceed in the contextual framework \cite{GR3}- \cite{GR1}. However, they understood contextuality 
differently from me and from the conventional viewpoint (as joint measurement with a compatible observable);
for them, context is not simply a complex of experimental physical conditions, see \cite{GR3}- \cite{GR1} for details.  Thus in this note we operate with three different types of 
contextuality.}  is trivially and commonly acceptable. In fact, this is not the case. And even if it were the case, it is very important to emphasize realism of QM. The main problem 
is wide spread of identifying of  ontic and epistemic realisms. As was discussed, the impossibility of this identification was interpreted as the impossibility to keep ontic realism (at least 
locally).
However, what is the most surprising this rejection of (local) ontic realism is often represented as total rejection of realism in QM.

\subsection{V\"axj\"o interpretation of quantum mechanics}

The V\"axj\"o interpretation (VI) of QM is the realist statistical local and contextual interpretation of QM \cite{VI1}-\cite{VI3}. 
We shall briefly explain the meanings assigned to these terms. 
\begin{itemize}

\item The easiest for explanation is ``statistical''. 
By VI probabilities are interpreted statistically, i.e., they are related to ensembles of systems and not individual 
systems. 

\item VI is local, simply no need even to rise the issue of nonlocality -
 in complete agreement with Einstein who mentioned nonlocality is an absurd  alternative to realism. 

\item VI considers both realisms: ``subquantum  realism'' (cf. Schr\"odinger, Einstein, De Broglie, Lochak, Atmanspacher and Primas) and ``contextual realism''
(cf. Bohr, Auff\`eves and Grangier). The first one is more challenging. We all still suffer of the lack of imagination... 
The second one was clearly presented
in section \ref{QR}.  

\item VI is contextual, as any interpretation referring to two descriptive levels, because any epistemic model is contextual as taking into account all experimental 
arrangement.  

\end{itemize}

\section{Future}

Recent successful tests for Bell's inequality can be interpreted in various ways, with two extremes, CV1-CV3 and NC1-NCV3, with numerous intermediate  interpretations.
 We can say that experimenters did their job and excellently. However, the loophole free Bell tests should not be interpreted as finalizing the century long discussion
about the possibility to construct a theoretical subquantum model coupled to the quantum observational model. 

My interpretation is that the successful realization of Bell's test confirmed that
\begin{itemize}
\item  We have to distinguish sharply  two descriptive levels, theoretical and observational (ontic and epistemic).
\item  The correspondence between them can be very tricky (so, not so straightforward as in the Bell conjecture {\bf O=E}).\footnote{Cf., e.g.,  with prequantum classical statistical field theory \cite{Beyond}. 
From my viewpoint this model reproduce all probabilistic predictions of QM for one and bipartite systems; for tripartite systems the representation of entanglement becomes extremely 
technical and hardly interpretable; for quadripartite systems I was not able to resolve technical difficulties.} 
\item Contextuality plays the crucial role in coupling     
of the two descriptive levels. 
\end{itemize}

Thus people looking for new insights on QM have to work hardly to find the proper subquantum theoretical description by relaxing constraints on the
rules of correspondence between the  two descriptive levels.

The final loophole free test says goodbye to the last hopes to identify the observational and ontic descriptions.  One has to look for more complex maps
\begin{equation}
\label{ZZTT3}
f: \rm{SUBQM}\to \rm{QM}.
\end{equation}

This  modeling has to be physically constructive, i.e., from the very beginning one should put efforts not to beat some no-go theorem(s), but to 
create a physically meaningful theoretical model\footnote{It may be surprising to hear this statement from a pure mathematician. However, in contrast to physicists who 
are so excited by the ability of mathematics to model everything in nature, I recognize that sometimes this ability looks like the ability of a prostitution to serve all men's
desires. Thus a purely mathematical subquantum model is practically valueless.}$\rm{SUBQM}$ endowed  with a physically meaningful map $f.$ A {\it partial} experimental verification 
is a part of this great project which successful realization would finally resolve (as I believe!!!) the puzzle of QM.   

Personally I still belief \cite{Beyond} in the possibility to construct  a subquantum model of the classical random field type,
cf.  Einstein and Infeld \cite{EI},  with nonlinear subquantum dynamics. This is also the main message of De Broglie. Although, in contrast to Einstein and Infeld and the author of this note,
he considered not the ordinary say classical electromagnetic field, but the so-called guiding field, this field is also physical. Thus one may hope 
to measure it, in some way.  Coming back to the Einstein-Infeld hypothesis about classical subquantum field, we can say that,   for a moment, experimentalists can try to 
proceed to new measurement technologies which would provide a possibility to measure characteristics of classical fields
for quantum systems, e.g., the polarization vector.   

\section*{Acknowledgments} I was lucky to discuss the Bell test and its possible interpretations with practically all world's leading experts in quantum foundations, 
those who firmly support the CV1-CV3 position as well as those who play with various variants of NCV1-NCV2. I would like thank all them for 
discussions and insights.  I am especially thankful to A. Zeilinger who expressed in lovely Viennese-style discussions  the firm anti-realist interpretation and K. Hess who expressed 
in the same lovely Viennese manner the firm realist position. It was very important for me that both are highly qualified {\it physicists,} that they do not simply 
play with mathematical formulas, but there is a lot of physical intuition behind their positions. And, of course, I was lucky to meet personally A. N. Kolmogorov and speak 
with him about foundations of CP (well, just 15 min conversation of the old and sick great man with the PhD-student who wrote a paper about a 
``non-Komogorov probability model'').  

The first test of this note was performed at March 18, 2016, at the Quantum Seminar of Stockholm University; I would like to thank all participants of this seminar and especially 
I. Bengtsson and  M. A. M. Versteegh for constructive critique. For writing of this paper the email correspondence with H. Atmanspacher and M.  Kupczynski was very stimulating.

This research is a part of the project ``Mathematical Modeling of Complex Hierarchic Systems" of the Faculty of Technology of Linnaeus University; 
it is also supported by Quantum Bio-Informatic Center of Tokyo University of Science.

\end{document}